\documentclass[a4paper,11pt]{article}
\pdfoutput=1
\usepackage{graphicx,mathptmx,helvet,courier,amsmath,mike}

\begin{document}

\title{A model of dissociated cortical tissue}

\author{Michael Stiber \and Fumitaka Kawasaki\\
Computing \& Software Systems \\
University of Washington, Bothell \\
Bothell, WA, 98011-8246 USA \\
stiber@u.washington.edu\\
fumik@u.washington.edu \\[2\parskip]
Dongming Xu \\
Linear Technology \\
15100 Weston Parkway Suite 202 \\
Cary, NC 27513 USA \\
dxu@linear.com}

\date{}

% \maketitle
% Save space by making the titles myself
\begin{center}
\textbf{\Large A MODEL OF DISSOCIATED CORTICAL TISSUE\footnote{A more
    complete version of this paper is available at http://faculty.washingon.edu/stiber/Public/NC07.pdf.}}\\[0.15in]

\vskip 16 pt

\begin{tabular}{p{3in}p{3in}}
\centering
\textit{Michael Stiber}\footnotemark \qquad \textit{Fumitaka Kawasaki}\\
Computing \& Software Systems Program \\
University of Washington, Bothell \\
Bothell, WA, 98011-8246 USA \\
stiber@u.washington.edu -- http://faculty.washington.edu/stiber \\
fumik@u.washington.edu &
\centering
\textit{Dongming Xu} \\
Linear Technology \\
15100 Weston Parkway Suite 202 \\
Cary, NC 27513 USA \\
dxu@linear.com
\end{tabular}
\end{center}

\footnotetext{To whom correspondence should be sent.}

\section*{ABSTRACT}

A powerful experimental approach for investigating computation in
networks of biological neurons is the use of cultured dissociated
cortical cells grown into networks on a multi-electrode array. Such
preparations allow investigation of network development, activity,
plasticity, responses to stimuli, and the effects of pharmacological
agents. They also exhibit whole-culture pathological bursting;
understanding the mechanisms that underlie this could allow creation
of more useful cell cultures and possibly have medical
applications~\cite{vanpelt-etal04,wagenaar-etal05a}.

\vskip 11pt

This paper presents preliminary results of a computational study of
the interplay of individual neuron activity, cell culture development,
and the network behavior.  We investigate whether bursting can occur
in an initially unconnected ``network'' that develops connections
according to an experimentally-verified model of cell culture
connectivity growth.

\paragraph{Neuron Model}
An integrate-and-fire neuron model with dynamical synapses that
exhibit activity-dependent facilitation and depression was
used~\cite{tsodyks-etal98,tsodyks-etal00,maass-etal02}.  It includes
synaptic, constant, and noise currents, with a reset of its membrane
voltage to a fixed value upon exceeding threshold and generating a
spike and a fixed absolute refractory period thereafter.

\vskip 11pt

The synapse has four state variables: three that govern the fraction
of synaptic resources in particular states --- $x$ (recovered state),
$y$ (active state), and $z$ (inactive state) --- and one, $u$, that
represents synaptic efficiency (see the more complete
paper\footnotemark[1] for detailed equations). See~\cite{maass-etal02}
for parameter values used.

\paragraph{Network Model}
Simulations were conducted by constructing networks with model neurons
on a rectangular grid. Connectivity was determined by incorporating a
model of cortical cell culture connectivity
development~\cite{vanooyen-etal95} that model's neurite outgrowth as a
radius of connectivity that changes at a rate inversely proportional
to a sigmoidal function of cell firing rate:
\begin{align}
  \fderiv{R_i}{t} &= \rho G(F_i) \label{eq:outgrowth} \\
  G(F_i) &= 1 - \frac{2}{1 + \exp((\epsilon - F_i )/\beta)}
\end{align}
where $R_i$ is the radius of connectivity of neuron $i$, $F_i$ is
neuron $i$'s normalized firing rate, $\rho$ is a rate constant,
$\epsilon$ is a constant that sets the ``null point'' for outgrowth
(the normalized \emph{target} firing rate that causes no outgrowth or
retraction), and $\beta$ determines the slope of $G(\cdot)$. One of
the parameters varied in these simulations was $\epsilon$.

\vskip 11pt

Synaptic strengths $W_i$ were computed for all pairs of neurons that
had overlapping connectivity regions as the area of their circles'
overlap. The bulk of the neurons in the network were excitatory; a
small number were chosen to be inhibitory. Similarly, most neurons
were not spontaneously active, but a few had lowered firing thresholds
to produce spontaneous firing at a rate of between 0.02 and 6
spikes/sec. To produce more consistent simulation results, a set of
standardized layouts was chosen to maximize spacing among inhibitory
and spontaneously active cells and reduce edge effects.  The fraction
of cells that were inhibitory was the other simulation parameter
varied.

\paragraph{Computer Implementation}
We used CSIM (A Neural Circuit SIMulator) version 1.1 for the
simulations. The original code was pared down to a small core that was
linked to a stand-alone C++ program to run on Linux, Windows, and
Macintosh computers. Generally speaking, each simulation took between
two and 20 hours on computers with 2--3GHz microprocessors.

\vskip 11pt

Simulations consisted of networks of 100 neurons in a 10x10
arrangement simulated for 30,000--60,000 seconds. While the rate of
neurite outgrowth was greatly sped up compared to the living
preparation, numerical investigation indicated that this did not
introduce instability in the simulation.

\paragraph{Analysis Methods}
To examine global behavior, average neuron firing rate and
\emph{burstiness index} (BI)~\cite{wagenaar-etal05a} were calculated
and plotted versus the two simulation parameters. BI was computed by
first calculating a spike count vs. time histogram for the entire
network during the last 5,000sec of the simulation. The fraction,
$f_{15}$, of the total number of spikes contained by the 15\% most
populous bins was then normalized to produce the burstiness index, BI,
as $\mathrm{BI} = (f_{15} - 0.15)/0.85$.  Detailed examination of
single simulations involved plots of neurons' connectivity radii and
firing rate versus time.

\begin{figure}
  \centering
  \begin{tabular}{rlrl}
    \raisebox{1.75in}{\textsf{\large A}} &
    \includegraphics[width=2.5in]{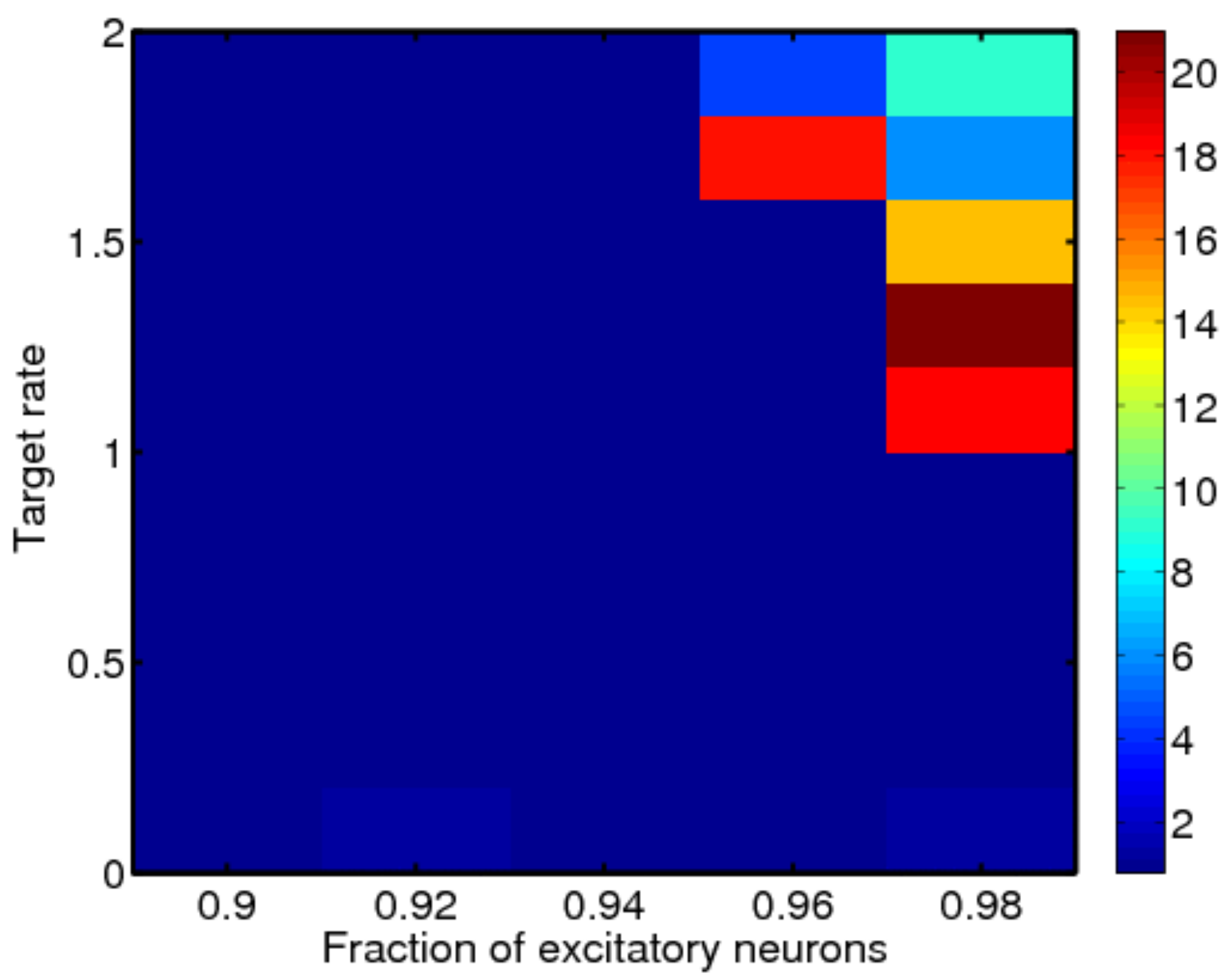} &
    \raisebox{1.75in}{\textsf{\large B}} &
    \includegraphics[width=2.5in]{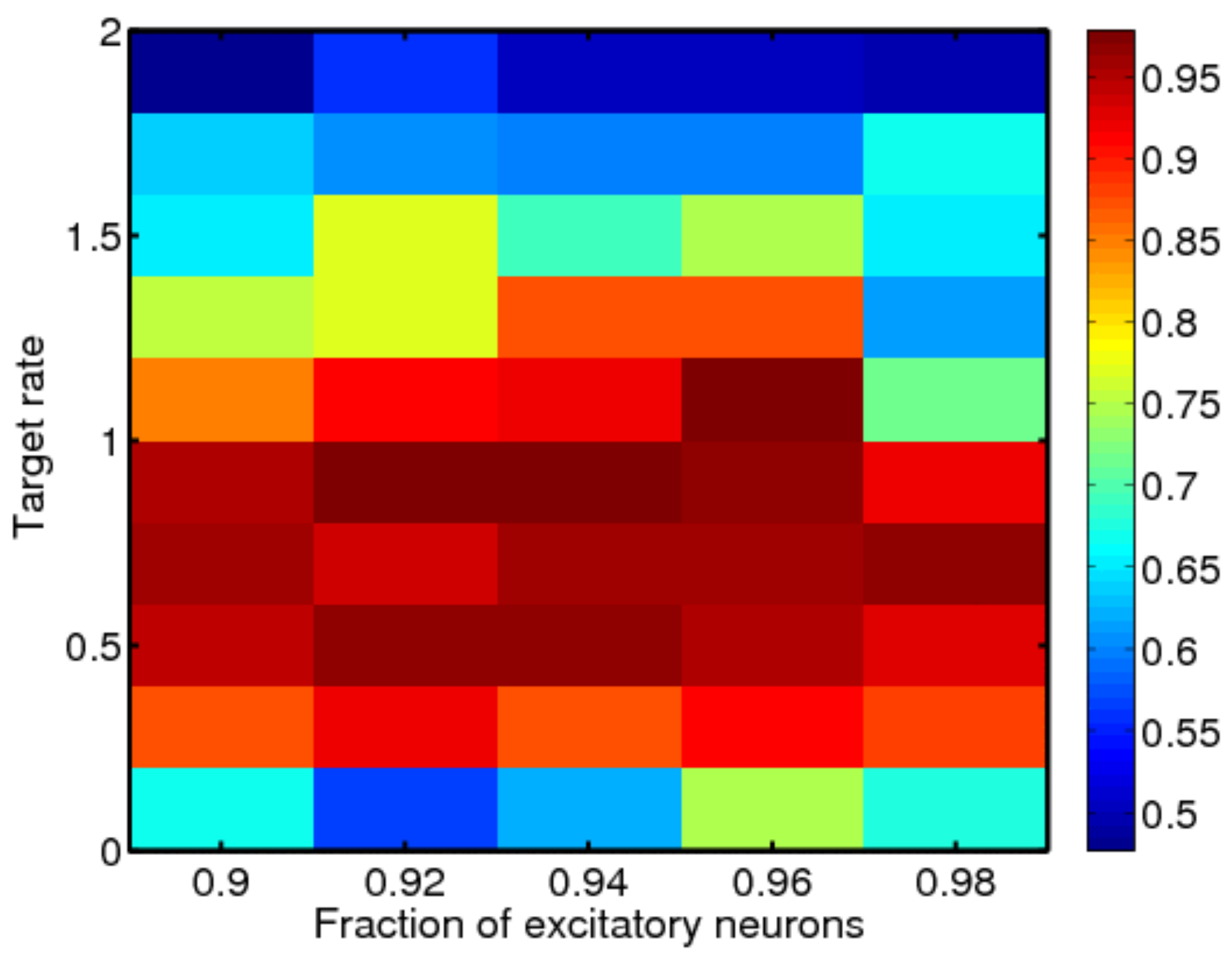}
  \end{tabular}
  \caption{Firing rates normalized relative to target rate (A);
    burstiness index (B). Both computed for 25,000-30,000sec and
    plotted versus the two simulation parameters.\label{fg:rate-BI}}
\end{figure}

\paragraph{Results}
An initial set of 50 simulations was performed with target rates in
the range 0.1--1.9 (inclusive, in 10 steps) and fraction of excitatory
neurons 0.9--0.98 (inclusive, in five steps).
Figure~\ref{fg:rate-BI}(A) shows normalized firing rates for the final
5,000sec of each simulation. Only the simulations with higher target
rates and fewer inhibitory neurons showed the great increase in firing
rate that might be associated with bursting. Longer (60,000sec)
simulations exhibited some bursting type of behaviors with the
fraction of bursting as low as 0.9 for the higher target rates.
Figure~\ref{fg:rate-BI}(B) indicates that the fraction of excitatory
neurons in this range has only a modest effect on burstiness index
(for 30,000sec simulations), and that moderate target rates produce
the highest BI values. This apparent conflict with the previous
observation of bursting at higher target rates can be explained by
examining the detailed behavior of individual simulations.

\begin{figure}
  \centering
  \begin{tabular}{ccc}
    \includegraphics[width=1.9in]{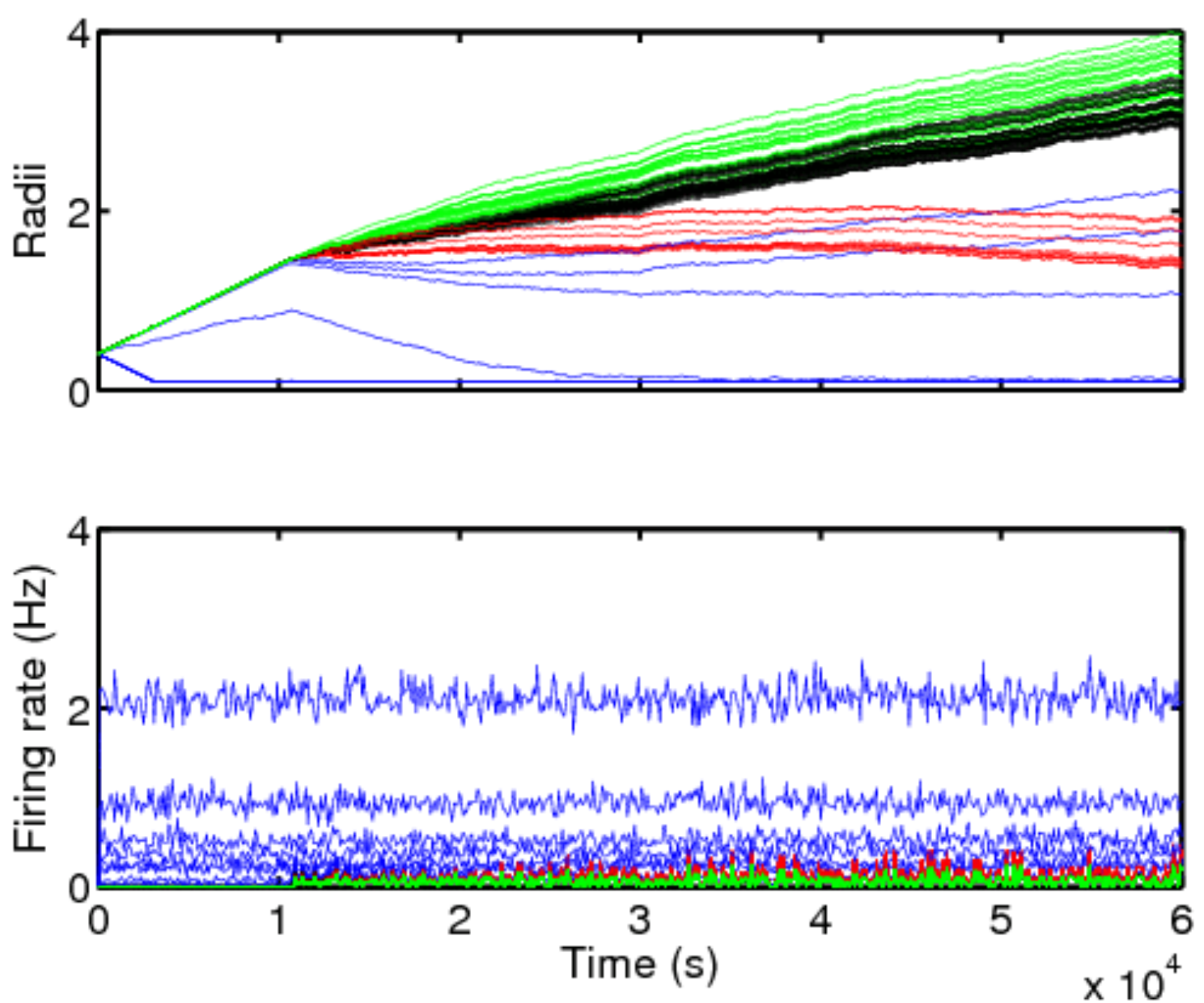} &
    \includegraphics[width=1.9in]{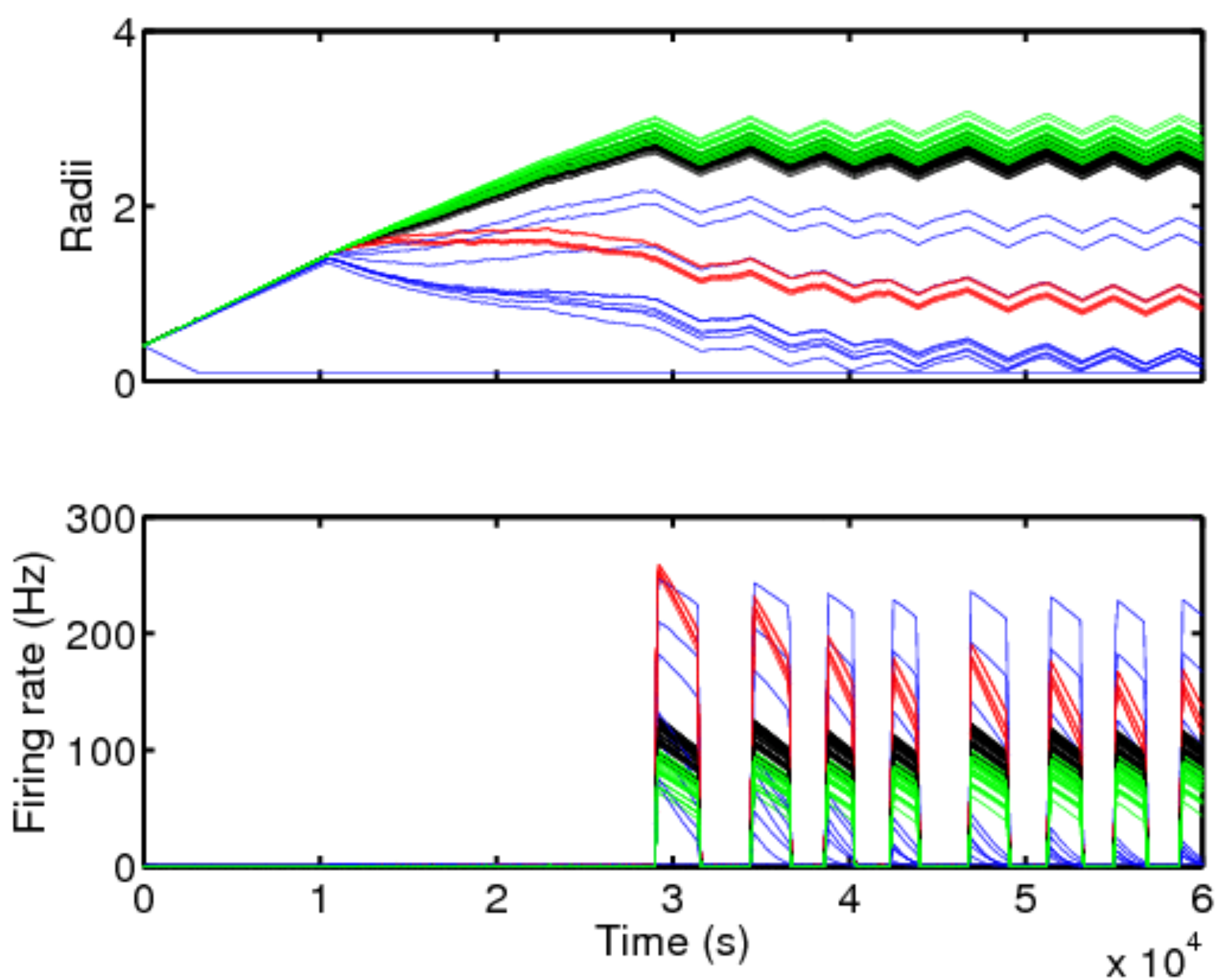} &
    \includegraphics[width=1.9in]{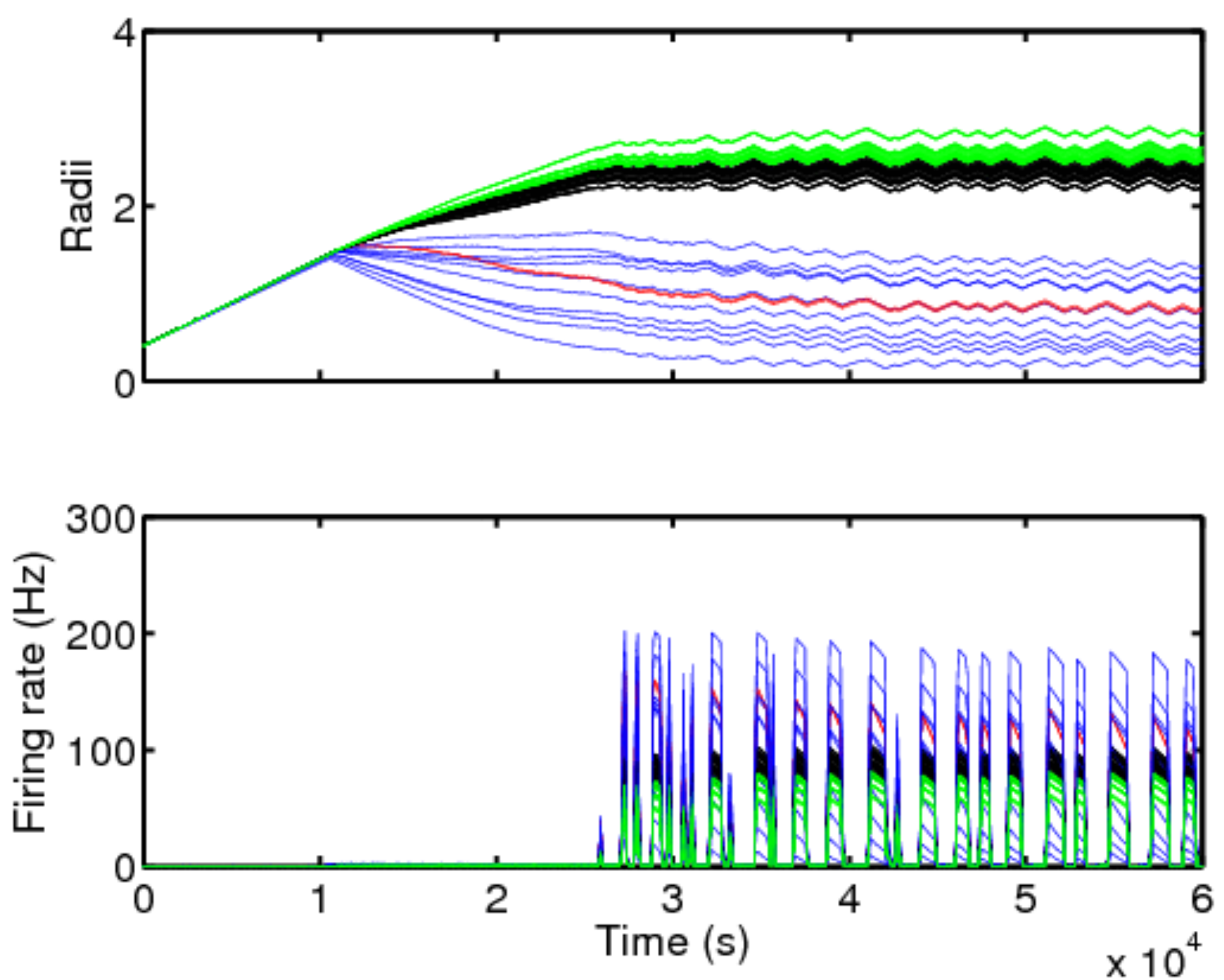} 
  \end{tabular}
  \caption{Detailed simulation results for simulations with parameters
    (target rate, fraction excitatory cells) of (0.1, 0.9) (left),
    (0.9, 0.94) (middle), and (1.9, 0.98) (right). Data for edge
    neurons are green, non-edge (and non-inhibitory, non-spontaneously
    active) neurons are black, inhibitory neurons are red, and
    spontaneously active neurons are blue.\label{fg:detail}}
\end{figure}

Figure~\ref{fg:detail} shows detailed information for simulations with
three sets of parameters: (target rate, fraction excitatory cells) =
(0.1, 0.9), (0.9, 0.94), and (1.9, 0.98). These include the parameter
extremes and a central value and both bursting and non-bursting
activity. In these cases, 60,000sec simulations were performed. Note
that some of the simulations that weren't bursting at 30,000sec were
bursting shortly thereafter, as evidenced by the (0.9, 0.94) one.

\vskip 11pt

Nevertheless, this confirms that the low BI values for low target
rates correspond to non-bursting behaviors (and that the connectivity
radii had not stabilized for cells that were not inhibitory or
spontaneously active). Low BI values for high target rates were a
possible result of the very broad or frequent bursts. For the bursting
behaviors, connectivity radii have stabilized, excepting small
variations during bursting.  In all simulations, connectivity radii
for edge neurons are larger than others, inhibitory neurons had
moderate radii, while spontaneously active neurons had a wide range of
different connectivities, likely due to the variability in their
firing thresholds.

\vskip 11pt

In either bursting or non-bursting behaviors, spontaneously active
neurons tended to be the most active. This is not surprising, as their
lowered thresholds would make them more excitable. The next higher
firing rates belonged to the inhibitory cells, then non-edge cells,
then edge cells.

\begin{figure}
  \centering
  \begin{tabular}{ccc}
    \includegraphics[width=1.9in]{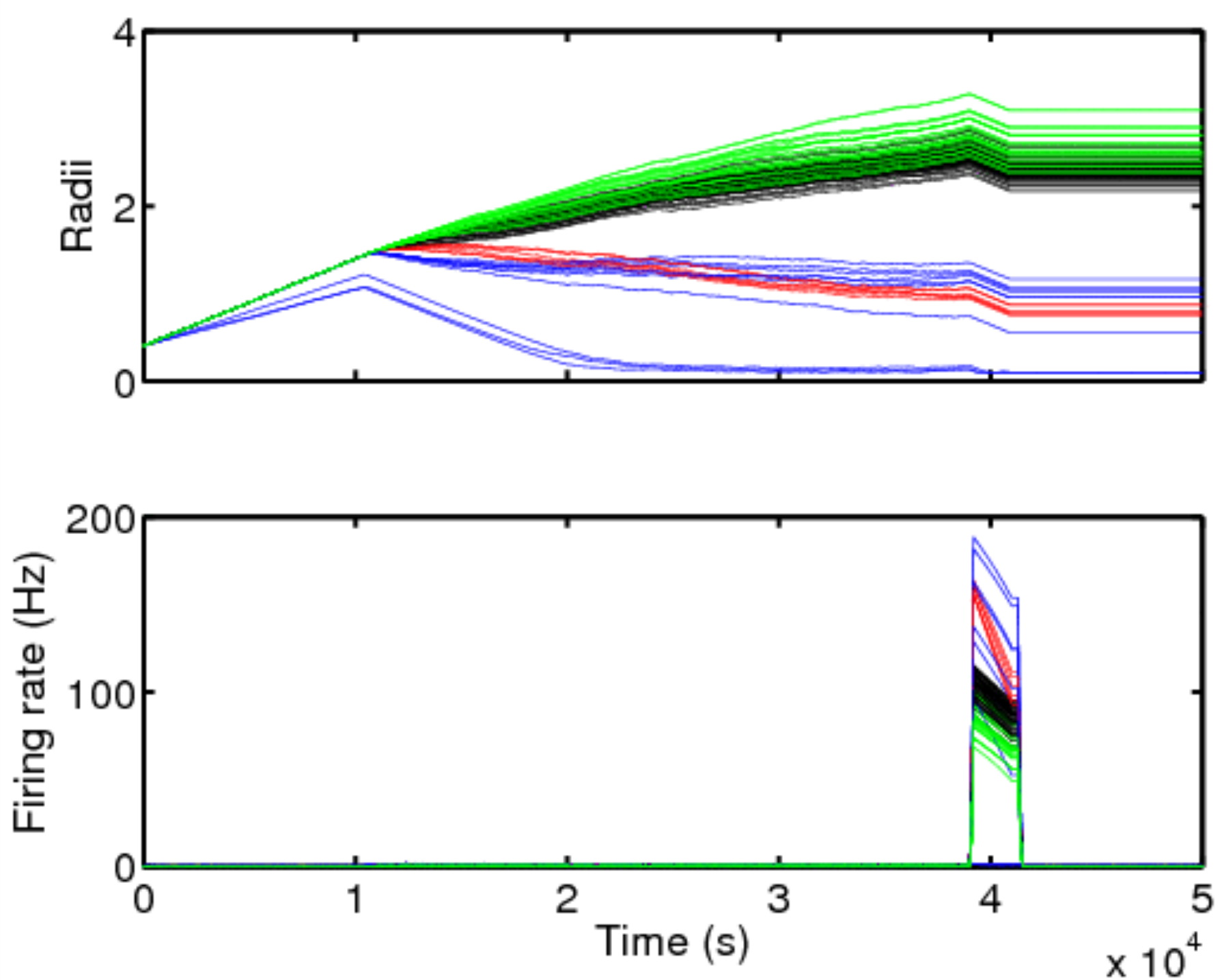} &
    \includegraphics[width=1.9in]{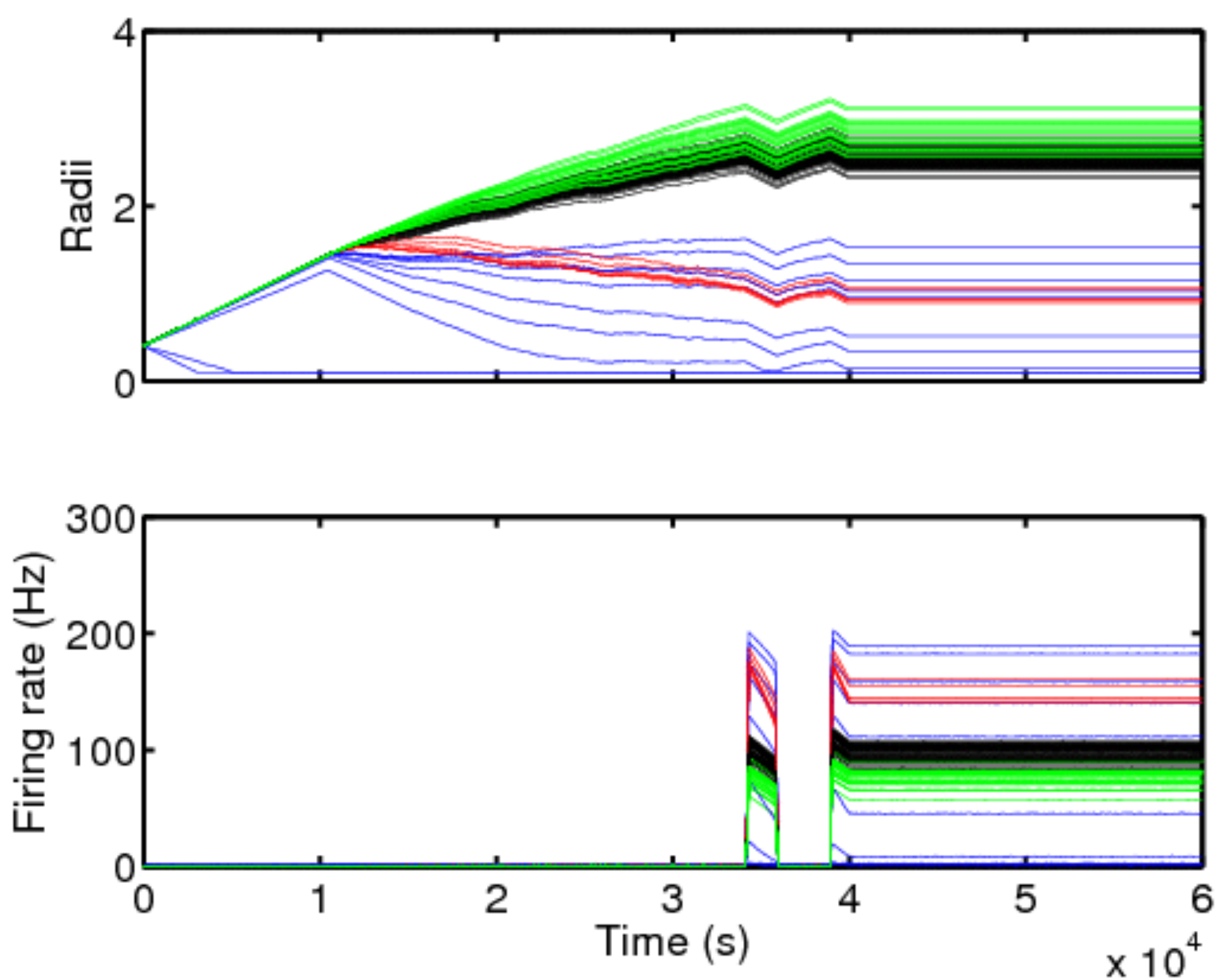} &
    \includegraphics[width=1.9in]{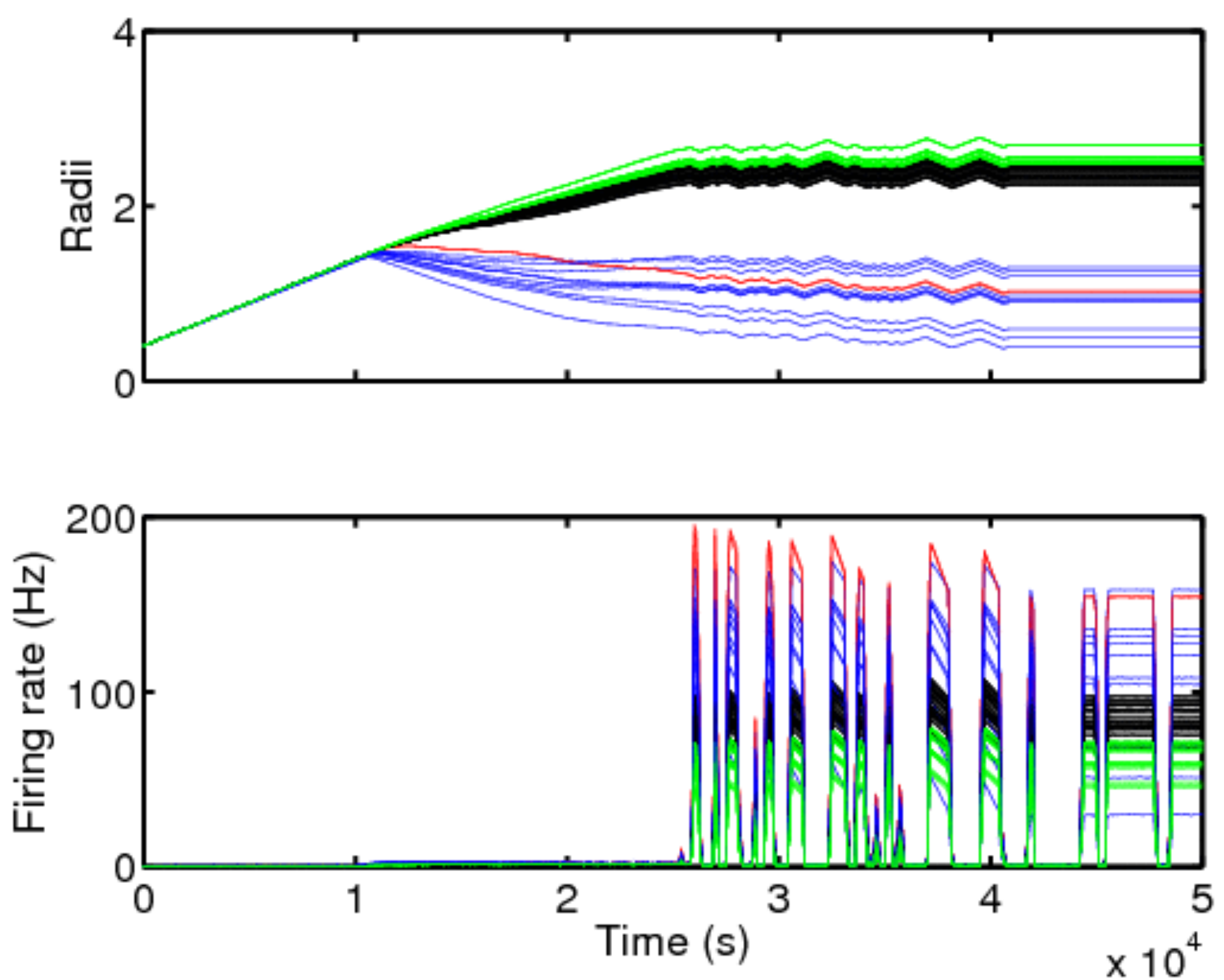}
  \end{tabular}
  \caption{Effects of neurite outgrowth on bursting for simulations
    with parameters (target rate, fraction excitatory cells) of (0.9,
    0.94) (left, middle) and (1.9, 0.98) (right). Growth was
    ``frozen'' either during (left, right) or between (middle)
    bursts.\label{fg:growth-burst}}
\end{figure}

In the results in figure~\ref{fg:detail}, it seems possible that the
mechanism for burst initiation and termination is the variation in
connectivity. Figure~\ref{fg:growth-burst} presents simulations in
which connectivity was fixed either during bursts (left, middle) or in
between bursts (right). For the lower target rate/lower fraction of
excitatory cells simulation, it does indeed seem that bursting is
controlled by connectivity. However, for a higher rate/higher fraction
of excitatory cells, bursting can continue even in the absence of
variation in connectivity (this simulation also produced bursts when
growth was stopped during a burst).

\paragraph{Discussion}
Bursting occurred with sufficiently small inhibition and high target
firing rate. One might expect the latter to produce greater
connectivity for every neuron, which in turn would be the mechanism
for whole-culture bursting. However, in the low-target-rate,
non-bursting simulations, such as figure~\ref{fg:detail}(left),
non-inhibitory, non-spontaneously active cells grow large connectivity
radii. Spontaneously active cells, on the other hand, tend to have
large connectivity radii in the bursting simulations. Presumably,
lowering these cells autonomous firing rate would result in bursting
at lower target rates.

\vskip 11pt

In previous investigations of bursting with randomly connected
networks~\cite{tsodyks-etal00}, the model synapses' depression and
facilitation were, neglecting the influence of noise, the mechanism
underlying burst initiation and termination. Our preliminary results
indicate that this is possibly the case under certain circumstances,
but not all. For some regions of parameter space, it may be the case
that the mechanism is a hysteresis effect involving changing
connectivity radii.  There are a number of possible reasons for this
difference:
\begin{itemize}
\item A number of parameters were set arbitrarily or not fully
  explored. These include fraction of excitatory cells, scaling of
  synaptic weights from area of connectivity overlap, and no
  differential scaling based on type of synapse (i.e., inhibitory
  vs. excitatory).

\item In the current simulations, only spontaneously active cells had
  any parameter variability; all other cells of a given type
  (inhibitory or excitatory) had identical parameters.

\item The overriding issue here is likely the small network size. Edge
  effects were great (edge neurons' connectivity radii were always the
  greatest of all neurons and 36/100 of the cells were edge neurons)
  and there were small numbers of inhibitory and spontaneously active
  cells. The final networks were almost completely
  connected. Increasing network size to, say, 100x100, could have
  little impact on final connectivity radii but with each neuron
  having connections to less than 10\% of the network.

  Increasing network size will have computational consequences that
  must be addressed: in its current form, a 60,000sec simulation of a
  100x100 network would take at least 2,000 hours (83 days).
\end{itemize}

There are also fundamental differences between the connectivity
patterns generated by this model (perhaps most similar to radial basis
functions) and many other models of cortex or recurrent networks (in
which either network topology is irrelevant or a power law-type
distribution is used that produces mostly local connections with a few
long-range ones). It will be instructive to investigate the detailed
correlation structure of inter- and intra-burst neuron firing.

\vskip 11pt

\textbf{\textit{Keywords:}} cortical cultures, network development, bursting.

\end{document}